# Quasi liquid layer-pressure asymmetrical model for the motion of of a curling rock on ice surface

Yuze Hao[1] and Yueqi Wang[2]

In this paper, we present a new model based on Quasi liquid layer to explain why the direction of lateral motion of the culing rock on ice surface is opposite to the other material surface. As we know, under the action of inertial force, the pressure on the ice surface in front of curling is greater than that on the back. So we assert that the firction coefficientin front of curling is lower than that on the bank under different pressure. In order to explain the pressure impact on friction coefficient, we qualitatively account for the reason why the coefficient of friction increases under the pressure and approximately calculated the relationship between the pressure and the thickness of the quasi_liquid layer on the ice surface. Then we calculate the function expression between temperature , pressure and the firction coefficient by the function between temperature and friction coefficient.

1. **Introduction:**

In the sport of curling, curling rocks is a granite with a steel contact annulus, when an athlete throws a curling rock, the curling rock rotates rather than moves in parallel, the trajectory of curling rock is curved rather than a straight line. Similar curling rocks sliding on dry surfaces have lateral motions in the opposite direction.Scientists is puzzled about the cause of the lareral motion of curling rocks on ice surface in the opposite direction to other materal surface. Many scientists presented some models for explan this puzzle . Some previous work are listed below:

a) Johnson presented a physical model that was based on an asymmetry in the firction around the contact annulus .

b) Mark shegelski presented a model that took into account the kinetic melting of the ice to give a thin liquid film nested between portions of the contact annulus and the underlying solid ice [1]. Later,in the paper [2],prevented the semi-phenomenological description of flooded ice and pebbled ice(consists of mm-sized ice mounds called "pebbles").But not all ice is flooded or peddled.

c) Harald Nyberg presented a model that the contact annulus will be scrach the ice[3] .The scraches will change the trajectory of the curling rocks.But paper [2]mentioned the dependence of the net curl on the initial angular speed is weak,which is contradictory . Simlarly , in the paper [4]can easily find contradictory .

d) Harrington presented that the firction at the bank of the running band is greater than that at the frount of the running band at 1924 [5]and 1930 [6].

e) Penner presented the theoretical and experimental results for sliding rotating cylindrical shells and curling rocks.

f) Marmo and Blackford describe a numerical model of curling stone dynamics that is based on

[1] Corresponding author ,E-mail address:haoyuze2021@outlook.com ORCID:0000-0001-7979-4242
[2] As Co-first author

the firction coefficient derived by stiffer[7] .

g) Jiro Murata proposed that the change of instantaneous rotation center leads to abnormal curl of curling rock at 2022 [8], and a large amount of experimental date have been obtained.

Although pebbled may be a reason for the abnormal trajectory , but it does not explain the normal situation .

Scraches-guide model can also explain the phenomenon to a certain extent , but scratches may not produce such a large steering force on a 20kg curling rock. At the same time, scratches-guide model may cause abnormal curl on the surface of other materials of curling rock, not just on the ice surface.

Almost all models previously presented for motion are based on heating the ice and asymmetrical of the speed but not consider the pressure assymmetry. But when the temperature is greater than - 7℃, the friction coefficient increases with the increase of temperature [11], at that time heating the ice cann't explain the abnormal phenomenon of curling rock. The speed assymmmetry model is mainly left-right asymmetry and front-bank asymmetry. In a publication[9] we can find that left-right asymmetrial are probably fault, so we probably need to consider the difference pressure between front and bank.

2. **Introduction to the new model**

We propose a new front-back asymmetrical model based on quasi liquid layer . We asserts that the friction coefficient is positively related to the thickness of the quasi-liquid layer in a certain range (hence called the quasi liquid layer–pressure asymmetrical model). The thickness of quasi liquid layer may become thicker with the increase of pressure. We approximately calculated the relationship between the pressure and the thickness of the quasi liquid layer on the ice surface by calculating the average amplitude of surface molecules. Then we calculate the function expression between pressure, temperature and firction coefficient, namely μ（p,T）based on the function between temperature and steel-ice firction coefficient. By using $\mu(p, T)$, we can find that the effect of pressure on the friction coefficient is similar to that of temperature, and we can calculate the monotone interval $\mu(p, T_0)$, where $T_0$ is a constant, and heating the ice will reduce the friction coefficient under -7℃. When the temperature is greater than - 7℃, the steel-ice friction coefficient increases with the increase of temperature [11], which needs to be explained by pressure asymmetry.

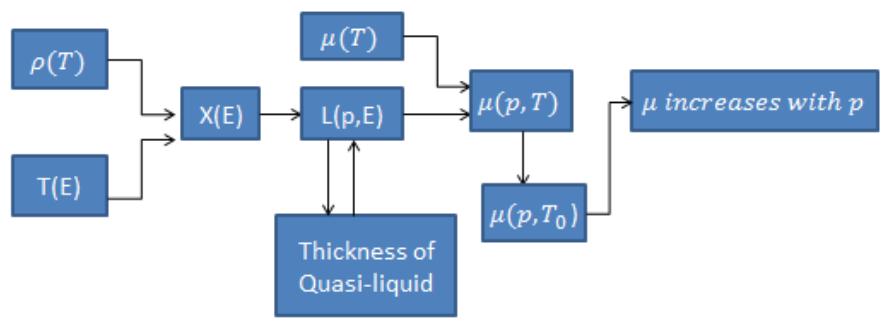

Figure 1：Model build process

Where L(p,E) is average amplitude of surface molecules in vertical direction.

3. **Pressure asymmetrical**

Due to the effect of inertial force, the pressure on ice surface before and after the speed of curling rock is asymmetric. In order to obtain the function of pressure at position θ, we divide the curling rock into infinitesimal elements at θ position, namely p(θ).

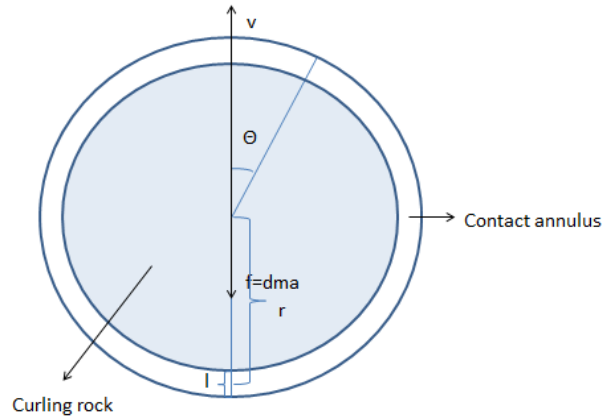

Figure 2: top view of curling rock

Where the thickness of contact annulus is l, curling rock has a radius of r. The acceleration of friction is a. We divide the curling rock into micro elements at θ position. Each micro element of curling rock receives the friction force of dma.

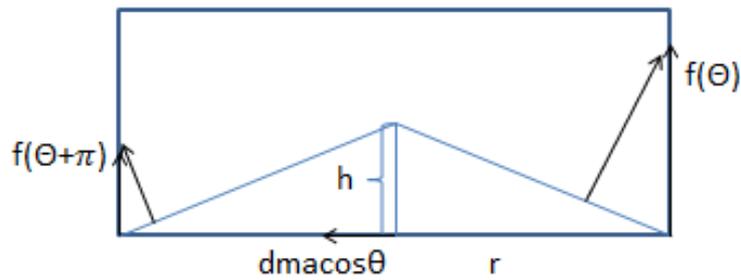

Figure 3: θ section of curling rock

According to Newton's third law: θ the moment of the preceding part of the tangent plane is equal to that of the preceding part θ moment of section.

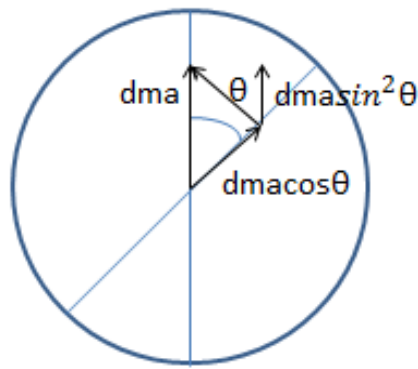

Figure 4: top view of free body diagram

Hence θ the moment of the latter part of the section is:

$$\int_\theta^{\frac{\pi}{2}} dmah\sin^2\theta = \int_\theta^{\frac{\pi}{2}} dmah(1 - \cos 2\theta) = \frac{\sin 2\theta}{4} dmah \quad (1)$$

According to the balance of couples condition in Figure 2, we can list the equation:

$$[f(\theta) - f(\theta + \pi)] \frac{r}{\sqrt{h^2 + r^2}} \sqrt{h^2 + r^2} = dmah\cos\theta + 2\int_\theta^{\frac{\pi}{2}} dmah\sin^2\theta \quad (2)$$

Divide both sides of equal sign by ds, $p=\frac{f}{s}$,

Hence:

$$[p(\theta) - p(\theta + \pi)]r = \frac{dm}{ds} ah(\cos\theta + \frac{\sin 2\theta}{2}) \qquad (3)$$

$$ds = lrd\theta \qquad (4)$$

Because:

$$\int_0^{\frac{\pi}{2}} [p(\theta) + p(\theta + \pi)] lrd\theta = mg \qquad (5)$$

Hence:

$$p(\theta) + p(\theta + \pi) = \frac{mg}{2\pi rl} = \frac{dmg}{ds} \qquad (6)$$

hence

$$\frac{dm}{ds} = \frac{m}{2\pi rl} \qquad (7)$$

Because of (3)(6), we can list:

$$P(\theta) - p(\theta + \pi) = \frac{mah}{2\pi lr^2}\left(\cos\theta + \frac{\sin 2\theta}{2}\right) \qquad (8)$$

By simultaneous equations (6) and (8), we can get:

$$P(\theta) = \frac{1}{2}[\frac{mg}{2\pi rl} + \frac{mah}{2\pi r^2 l}\left(\cos\theta + \frac{\sin 2\theta}{2}\right)] \qquad (9)$$

The mass of curling rock is about 20kg, the radius is about 15cm, the height is about 10cm, the thickness of contact annulus is about 1cm.

Hence the date of equation (9) is

| m(kg) | h(m) | l(m) | r(m) |
|---|---|---|---|
| 20 | 0.05 | 0.01 | 0.15 |

Hence equation (9) equals:

$$P(\theta, a) = \frac{1}{2}[21231 + 707a(\cos\theta + \frac{\sin 2\theta}{2})] \qquad (10)$$

According to equation(10), we can find the pressure in front of speed direction is greater than that on the back.

4. **Quasi-liquid layer**

In a publication [10] mention: The existence of a quasi-liquid layer on the surface of ice crystals at temperatures not far below 0°C is widely accepted on the basis of many experimental and theoretical studies. Quasi liquid layer causes low friction coefficient of ice, so the thickness of quasi liquid layer will naturally impact on the friction coefficient, pressure may cause changes in the thickness of the quasi liquid layer, which in turn impact on the friction coefficient.

5. **Date analysis**

In order to obtain the average amplitudes of surface molecules, we need to calculate the function expression between average distance of ice molecule and total energy under pressure in the direction perpendicular to the ice surface . To calculate the function expression , we need to obtain the equation between the density of ice and temperature, date of ice density is list below and we fit curves by using computer simulation.

| T(℃) | 0 | -5 | -10 | -15 | -20 | -25 | -30 | -35 | -40 |
|---|---|---|---|---|---|---|---|---|---|
| ρ (kg\$m^3$) | 0.9162 | 0.9175 | 0.9189 | 0.9194 | 0.9194 | 0.9196 | 0.9200 | 0.9204 | 0.9208 |
| T(°C) | -50 | -60 | -70 | -80 | -90 | -100 | | | |
| ρ (kg\$m^3$) | 0.9216 | 0.9224 | 0.9233 | 0.9241 | 0.9249 | 0.9257 | | | |

*Date from The Engineering Tool Box* [3]

Common fit curves listed below:

$$\rho = -8 \times 10^{-5}T + 0.9162 \quad R^2 = 0.9741 \tag{11}$$

$$\rho = -2 \times 10^{-7}T^2 - 0.0001T + 0.9172 \quad R^2 = 0.9776 \tag{12}$$

$$\rho = -9 \times 10^{-9}T^3 - 2 \times 10^{-6}T^2 - 0.0002T + 0.9168 \quad R^2 = 0.984 \tag{13}$$

$$\rho = -4 \times 10^{-10}T^4 - 9 \times 10^{-8}T^3 - 7 \times 10^{-6}T^2 - 0.0003T + 0.9164 \quad R^2 = 0.9925 \tag{14}$$

$$\rho = 0.9174 e^{-9 \times 10^{-5}T} \quad R^2 = 0.9737 \tag{15}$$

Considering the solvability and computational complexity, we choose (11) as the function ρ(T)

**6. Accounting qualitating and approximately calculated**

As we know:

$$dQ = C_{ice} m_{ice} dT \tag{16}$$

Where $C_{ice}$ is the specific heat capacity of ice, $m_{ice}$ is the mass of a single water molecule

$$\int dQ = \int C_{ice} m_{ice} dT \tag{17}$$

Hence the total energy E is:

$$E = Q = C_{ice} m_{ice} T \tag{18}$$

Hence

$$T = \frac{E}{C_{ice} m_{ice}} \tag{19}$$

Hence:

$$\rho(T) = \rho \left( \frac{E}{C_{ice} m_{ice}} \right) \tag{20}$$

so, ρ(E) equals:

$$-8 \times 10^{-5} \frac{E}{C_{ice} m_{ice}} + 0.9174 \tag{21}$$

Where $C_{ice} = 2.1 \times 10^3 J/(kg \cdot K)$, $m_{ice} = 2.99 \times 10^{-26} kg$, hence:

$$\rho(E) = 0.1592 \times 10^{18} E + 0.9174 \tag{22}$$

Because ice crystals are mainly ice-Ih structure, it is the cubic densest packing structure, according to the geometric relationship, consider average distance between adjacent molecules, hence:

$$\frac{4 m_{H_2O}}{\rho(E)} = \frac{3}{\sqrt{2}+1} x^2 \approx 1.917 x^3 \tag{23}$$

x is the approximate average distance between adjacent molecules

---

[3] The Engineering ToolBox, Ice-Thermal Properties.
https://www.engineeringtoolbox.com/ice-thermal-properties-d_576.html

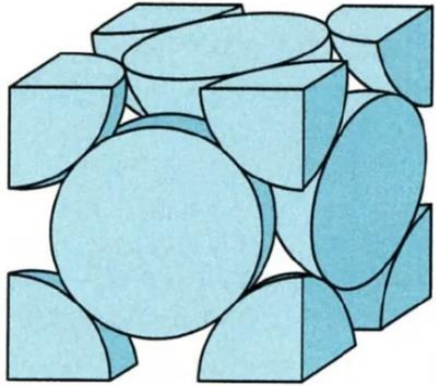 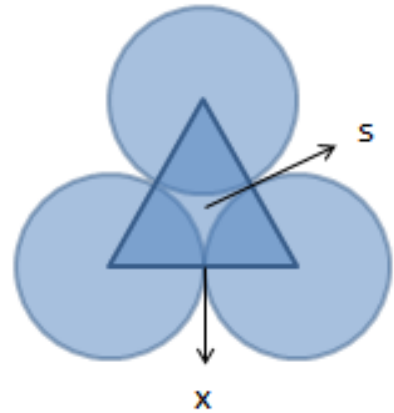

Figure 5: Cubic densest packing crystal structure    Figure6:Average area of surface molecules

The work done by presure is psx(s is the contact area of surface molecule)

The total energy under pressure is $E - psx$ ,s is the contact area of surface molecule ,the molecular diameter of water is $4\times 10^{-10}m$ ,hence:

$$s = 48\sqrt{3} \times 10^{-20} m^2 \tag{24}$$

so the average amplitude of ice surface molecules under pressure is:

$$L(p,E)=x(-E-psx)+x(E-psx) \tag{25}$$

X(E-psx)is one solution of the equation

$$1.917x^3\left[0.1592 \times 10^{18}(E - 48\sqrt{3}\times 10^{-20}px) + 0.9174\right] = 4m_{ice} \tag{26}$$

Hence,

$$psx^4 + \rho(E)x^3 - \frac{4m_{ice}}{1.917} = 0 \tag{27}$$

Judge by order of magnitude and Quadratic equation rooting formula, L(p,E) is:

$$L(p,E) =$$

$$\tfrac{1}{2}\sqrt{A_1 - A_2 + A_3 + \frac{2A_5}{4\sqrt{\tfrac{1}{2}A_1+A_4+A_3}}} + \tfrac{1}{2}\sqrt{-A_6 - A_7 + A_8 - \frac{2A_5}{4\sqrt{\tfrac{1}{2}A_6+A_9+A_8}}} + \tfrac{1}{2}\sqrt{\tfrac{1}{2}A_1 + A_2 + A_3} +$$

$$\tfrac{1}{2}\sqrt{\tfrac{1}{2}A_6 + A_7 + A_8} \tag{28}$$

$$A_1 = \frac{\rho^2(E)}{2(0.1592\times 10^{18}ps)^2}$$

$$A_2 = \frac{\sqrt[3]{2} - 0.1592\times 10^{18}ps\,\frac{4m_{ice}}{1.917}}{0.4776\times 10^{18}ps\,\sqrt[3]{27\rho^2(E)\frac{4m_{ice}}{1.917} + \sqrt{-4[(-0.1592\times 10^{18})ps\frac{4m_{ice}}{1.917}]^3 + [27\rho^2(E)\frac{4m_{ice}}{1.917}]^2}}}$$

$$A_3 = \frac{\sqrt[3]{-27\rho^2(E)\frac{4m_{ice}}{1.917} + \sqrt{-4(-0.1592\times 10^{18}ps\frac{4m_{ice}}{1.917})^3 + [27\rho^2(E)\frac{4m_{ice}}{1.917}]^2}}}{3\sqrt[3]{2}(0.1592\times 10^{18}ps)}$$

$$A_4 = \frac{3\sqrt{2}(-0.1592 \times 10^{18}ps)\frac{4m_{ice}}{1.917}}{0.4776 \times 10^{18}ps \sqrt[3]{-27\rho^2(E)\frac{4m_{ice}}{1.917} + \sqrt{4(0.1592 \times 10^{18}ps\frac{4m_{ice}}{1.917})^3 + [27\rho^2(E)\frac{4m_{ice}}{1.917}]^2}}}$$

$$A_5 = \frac{\rho^3(E)}{2(0.1592 \times 10^{18}ps)^2}$$

$$A_6 = \frac{\rho^2(-E)}{2(0.1592 \times 10^{18}ps)^2}$$

$$A_7 = \frac{\sqrt[3]{2} - 0.1592 \times 10^{18}ps\frac{4m_{ice}}{1.917}}{0.4776 \times 10^{18}ps \sqrt[3]{27\rho^2(-E)\frac{4m_{ice}}{1.917} + \sqrt{4[(0.1592 \times 10^{18})ps\frac{4m_{ice}}{1.917}]^3 + [27\rho^2(-E)\frac{4m_{ice}}{1.917}]^2}}}$$

$$A_8 = \frac{\sqrt[3]{-27\rho^2(-E)\frac{4m_{ice}}{1.917} + \sqrt{4(0.1592 \times 10^{18}ps\frac{4m_{ice}}{1.917})^3 + [27\rho^2(-E)\frac{4m_{ice}}{1.917}]^2}}}{3\sqrt[3]{2}(0.1592 \times 10^{18}ps)}$$

$$A_9 = \frac{3\sqrt{2}(-0.1592 \times 10^{18}ps)\frac{4m_{ice}}{1.917}}{0.4776 \times 10^{18}ps \sqrt[3]{-27\rho^2(-E)\frac{4m_{ice}}{1.917} + \sqrt{4(0.1592 \times 10^{18}ps\frac{4m_{ice}}{1.917})^3 + [27\rho^2(-E)\frac{4m_{ice}}{1.917}]^2}}}$$

Hence, the amplitude increases as the pressure increases. So as the pressure rises, the quasi liquid layer becomes thicker. We assert that the friction coefficient decreases with the increase of thickness within a certain range.

We can also find that heating and pressurization have the same effect on the thickness of the liquid layer.

7. **Firction coefficient and Quasi-liquid layer function**

In a publicantion [11], the experimental environment is standard atmospheric pressure $p_0$=101kpa we can find that the friction coefficient is the lowest at - 7 ℃(the total energy is 42.38× $10^{23}J$), when $L_0$=L(42.38× $10^{23}J$, 101000pa) the fiction coefficient is the lowest.

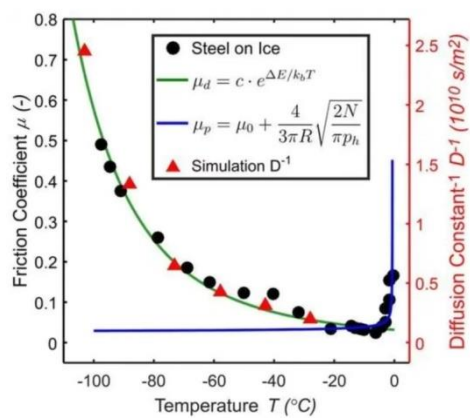

Figure7:function between temperature and friction coefficient, from [11]

$$Where\ \mu_0 \approx 0.03,\ \Delta E \approx 11.5\ kj.mol^{-1}$$

When T is greater than -7°C, friction coefficient increases is due to plastic ploughing, we only consider the impact of Quasi-liquid layer here, so we only discuss the function under -7°C.

Because:

$$\mu(T) = c \times e^{\Delta E/k_b T} \quad (29)$$

hence:

$$T(\mu) = \Delta E/k_b(ln\mu - lnc) \quad (30)$$

Hence the function expression between steel-ice firction coefficient and the thickness of Quasi-liquid layer is

$$L[101kpa, \Delta E/k_b(ln\mu - lnc)] \quad (31)$$

So the equiption between temperature, pressure and steel-ice firction coefficient μ(p,T) is

$$L(p,T) = L[101kpa, \Delta E/k_b(ln\mu - lnc)] \quad (32)$$

Because L(p,E) function is too complex, we will not list the equation expression in detail here. Through the equation of temperature, pressure and friction coefficient, we can easily get the equation of pressure and friction coefficient at constant temperature or changeing temperature (possible heat generated by firction).

## 8. Conclusion and outlook

One four main conclusions in this paper are:

i. Because the thickness of quasi liquid layer increases with the increase of pressure, the friction coefficient of ice decreases with the increase of pressure in a certain range.

ii. Pressure rise and temperature rise have a synergistic effect, so heating ice can also decrease the friction coefficient.

iii. The abnormal trajectory of curling rocks(-v× ω direction) may caused by quasi liquid layer.

iv. The equation of temperature, pressure and the friction coefficient is

$$L(p,T) = L[101kpa, \Delta E/k_b(ln\mu - lnc)]$$

In conclusion, the abnormal curl of curling rock is caused by the combined action of temperature and pressure. However, curling does not curl in the opposite direction on the surface of other materials not far below the melting point, which may be caused by hydrogen bonding. In the above calculation, the treatment of $\rho(T)$ takes into account the hydrogen bond

In the following research, we can quantitatively calculate the function of temperature pressure and friction coefficient.

It is hope that future work will be able to model the interaction between curling rock and the Quasi liquid film in a more complete manner, and express the relationship between fiction coefficient and pressure at constant temperature in a detailed way. It may cause impact on engineer，glaciology, etc.

## 9. Authors' Contributions

Yuze Hao proposed Quasi liquid layer-pressure asymmetrical model and specific algorithms and was responsible for writing the manuscript. Yueqi Wang is responsible for calculating complex equiptions in this paper.

## 10. ACKNOWLEDGEMENT

We would like to thank Dr.Alison Criscitiello from Canadian Ice Core Lab（CICL）for providing ice density date.